\documentclass[12pt]{article}
 \pdfoutput=1
\usepackage{amsfonts,amsthm,amsmath,amssymb,slashed}
\usepackage[textwidth = 500 pt, textheight = 630 pt]{geometry}
\usepackage{latexsym,amsfonts,amsmath,amssymb,theorem,dsfont,wasysym}
\usepackage{amssymb}
\usepackage{amsmath}
\usepackage{amstext}
\usepackage{graphicx,epsfig}
\usepackage{epsfig}
\usepackage{verbatim} 
\usepackage{fancyhdr}
\usepackage{fancybox}
\usepackage{color}
\usepackage{ulem,bbold}
\usepackage{enumitem}
\usepackage{subfigure}
\usepackage{bbm}
\usepackage{parskip}
\usepackage[numbers,sort&compress]{natbib}

\definecolor{MyDarkBlue}{rgb}{0.15,0.15,0.45}
\usepackage[linktocpage=true]{hyperref}
\hypersetup{
colorlinks=true,
citecolor=MyDarkBlue,
linkcolor=MyDarkBlue,
urlcolor=MyDarkBlue,
}

\usepackage[numbers,sort&compress]{natbib}
\usepackage{hypernat}

\def\beq{\begin{eqnarray}}
\def\eeq{\end{eqnarray}}

\def\({\left(}
\def\){\right)}

\newcommand{\be}{\begin{equation}}
\newcommand{\ee}{\end{equation}}
\newcommand{\la}{\langle}
\newcommand{\ra}{\rangle}
\def\ea{\end{eqnarray}}
\def\ba{\begin{eqnarray}}

\def\beq{\begin{eqnarray}}
\def\eeq{\end{eqnarray}}

\def\({\left(}
\def\){\right)}

\def\p{\partial}
\def\la{\langle}
\def\ra{\rangle}

\def\lsim{\mathrel{\rlap{\lower3pt\hbox{\hskip0pt$\sim$}}
     \raise1pt\hbox{$<$}}}         %less than or approx. symbol
\def\gsim{\mathrel{\rlap{\lower4pt\hbox{\hskip1pt$\sim$}}
     \raise1pt\hbox{$>$}}}         %greater than or approx. symbol

\def\lsim{\mathrel{\rlap{\lower3pt\hbox{\hskip0pt$\sim$}}
     \raise1pt\hbox{$<$}}}         %less than or approx. symbol
\def\gsim{\mathrel{\rlap{\lower4pt\hbox{\hskip1pt$\sim$}}
     \raise1pt\hbox{$>$}}}         %greater than or approx. symbol

\usepackage{amsmath}
\usepackage{amsfonts}
\usepackage{verbatim}
\usepackage{graphicx}
\usepackage{subfigure}
\usepackage{amssymb}
\usepackage[T1]{fontenc}
\usepackage{slashed}

\begin{document}

\renewcommand{\thefootnote}{\fnsymbol{footnote}}

\makeatletter
\@addtoreset{equation}{section}
\makeatother
\renewcommand{\theequation}{\thesection.\arabic{equation}}

\rightline{}
\rightline{}
   \vspace{1.8truecm}

%\begin{flushright}
% preprint nrs.
%\end{flushright}

\vspace{10pt}

%%%%%%%%%%%%%%%%%

\begin{center}
{\Large \bf{On the Initial State and Consistency Relations}}
\end{center} 
 \vspace{1truecm}
\thispagestyle{empty} \centerline{{\large  {Lasha Berezhiani and Justin Khoury}}
}

\vspace{1cm}
\centerline{{\it Center for Particle Cosmology, Department of Physics \& Astronomy, University of Pennsylvania,}}
 \centerline{{\it  209 South 33rd Street, Philadelphia, PA 19104}}
 
 \vspace{1cm}
 
\begin{abstract}

We study the effect of the initial state on the consistency conditions for adiabatic perturbations. In order to be consistent with the constraints
of General Relativity, the initial state must be diffeomorphism invariant. As a result, we show that initial
wavefunctional/density matrix has to satisfy a Slavnov-Taylor identity similar to that of the action. We then investigate the precise ways in which modified initial states can lead to violations of the
consistency relations. We find two independent sources of violations: $i)$ the state can include initial non-Gaussianities; $ii)$ even if the initial state is Gaussian, such as a
Bogoliubov state, the modified 2-point function can modify the $\vec{q}\rightarrow 0$ analyticity properties of the vertex functional 
and result in violations of the consistency relations.
\end{abstract}

\newpage
\setcounter{page}{1}

%\tableofcontents
\renewcommand{\thefootnote}{\arabic{footnote}}
\setcounter{footnote}{0}

\linespread{1.1}
\parskip 4pt

\section{Introduction}

Standard calculations of primordial correlation functions from inflation assume the Bunch-Davies~\cite{Bunch:1978yq} or adiabatic vacuum state.
Since quantum fluctuations originate from scales much smaller than the Hubble scale and therefore experience
a nearly flat background space-time, the argument goes, it is reasonable to assume they were born in a state which asymptotically approaches the flat-space vacuum.
This is by definition the Bunch-Davies vacuum. On the other hand, since not much is known about the high energy, pre-inflationary epoch, one is
justified to consider more general initial states/density matrices~\cite{Brandenberger:1999sw,Niemeyer:2000eh,Martin:2000xs,Brandenberger:2000wr,Kempf:2000ac,Niemeyer:2001qe,Easther:2001fz,Kempf:2001fa,Danielsson:2002kx,Shankaranarayanan:2002ax,Easther:2002xe,Kaloper:2002uj,Danielsson:2002qh,Kaloper:2002cs,Burgess:2002ub,Burgess:2003zw,Schalm:2004qk,Porrati:2004gz,Greene:2004np,Collins:2005nu}, as a way of parametrizing our ignorance about the conditions at the onset of inflation. These modifications have been argued to affect various late-time observables, in particular non-Gaussianities~\cite{Porrati:2004dm,Chen:2006nt,Holman:2007na,Meerburg:2009ys,Collins:2009pf,Meerburg:2009fi,Meerburg:2010ks,Agullo:2010ws,Ashoorioon:2010xg,Ganc:2011dy,Dey:2011mj,Chialva:2011hc,Kundu:2011sg,Dey:2012qp,Agarwal:2012mq,Flauger:2013hra,Aravind:2013lra,Ashoorioon:2013eia}. 

The goal of this work is to systematically study the impact of the initial state on the consistency relations for primordial perturbations.
Because of the few assumptions that go into them, the consistency relations offer a powerful probe of early-universe
physics. Specifically, their proof relies on: $i)$ a single scalar field; $ii)$ an attractor background (mode functions asymptote to a
constant at late times); $iii)$ the Bunch-Davies vacuum. We focus on $iii)$ and investigate to what extent the consistency relations can
be violated once we relax the Bunch-Davies assumption.

The consistency relations, derived recently as Ward identities~\cite{Hinterbichler:2013dpa,Berezhiani:2013ewa} for non-linearly realized global symmetries~\cite{Hinterbichler:2012nm},  
take the schematic form
\be
\lim_{\vec{q}\rightarrow 0} {\partial^n \over \partial q^n}
\left( \frac{ \langle \zeta_{\vec q} {\cal O}_{\vec{p}_1,\ldots,\vec{p}_N} \rangle}{P_\zeta (q)} + 
\frac{ \langle \gamma_{\vec q} {\cal O}_{\vec{p}_1,\ldots,\vec{p}_N} \rangle}{ P_\gamma (q)} \right)
\sim \sum_a {\partial^n \over \partial p^n_a}  \langle {\cal O}_{\vec{p}_1,\ldots,\vec{p}_N}
  \rangle  \, .
\label{schematicintro}
\ee
They constrain the $q^n$ behavior of an $N+1$-point correlation function with a soft scalar or tensor mode to a symmetry transformation on an $N$-point function. At lowest order ($n= 0$), one obtains Maldacena's consistency relations~\cite{Maldacena:2002vr,Creminelli:2004yq,Cheung:2007sv}, which relate the 3-point functions with a soft scalar or tensor to rescalings of the 2-point function. At the next order ($n= 1$) they are the linear-gradient consistency relations~\cite{Creminelli:2011rh,Creminelli:2012ed}. See~\cite{Assassi:2012zq,Goldberger:2013rsa,Schalm:2012pi,Bzowski:2012ih,Pimentel:2013gza,Collins:2014fwa} for related derivations of the $n= 0$, $n=1$ relations through Ward identities. The identities for $n\geq 2$, first discovered in~\cite{Hinterbichler:2013dpa}, partially constrain the soft limit of correlation functions in terms of lower-point functions. The $3\rightarrow 2$ relations have been checked explicitly up to and including $q^3$ order~\cite{Berezhiani:2014tda}.
 
The consistency relations have also been shown to derive from a single {\it master identity}, as a consequence of the Slavnov-Taylor identity for spatial diffeomorphisms~\cite{Berezhiani:2013ewa}. For soft 3-point functions, with the hard momenta given by scalar modes, the master identity takes the form
\beq
q^j\left(\frac{1}{3}\delta_{ij}\Gamma^{3{\rm d}\,\zeta\zeta\zeta}(\vec{q},\vec{p},-\vec{q}-\vec{p})+2 \Gamma_{ij}^{3{\rm d}\,\gamma\zeta\zeta}(\vec{q},\vec{p},-\vec{q}-\vec{p})\right)=q_i \Gamma_\zeta^{3{\rm d}}( p )-p_i\bigg( \Gamma_\zeta^{3{\rm d}}(|\vec{q}+\vec{p}|)-\Gamma_\zeta^{3{\rm d}}( p ) \bigg)\,,
\label{wardintro}
\eeq
where $\Gamma^{3{\rm d}\,\zeta\zeta\zeta}$ and $\Gamma^{3{\rm d}\,\gamma\zeta\zeta}$ are respectively the cubic vertex functions for 3 scalars, and for 2 scalars$-$1 tensor, while $\Gamma_\zeta$ is the inverse scalar propagator. This master identity is valid at any $q$ and therefore goes beyond the soft limit. By resuming to correlation functions and differentiating a number of times with respect to $q$, one recovers order by order all of the identities~\eqref{schematicintro}. This approach underscores the role of diffeomorphism invariance at the root of cosmological consistency relations and sharpens the precise assumptions necessary for their validity.

The consistency relations are {\it physical} statements and can be related to late-time observables, such as the CMB bispectrum~\cite{Creminelli:2011sq,Bartolo:2011wb,Pajer:2013ana} and the large-scale structure~\cite{Kehagias:2013yd,Peloso:2013zw,Creminelli:2013mca,Horn:2014rta}. Their violations can be tested observationally. For instance, had the Planck satellite detected a significant primordial local $f_{\rm NL}$, this would have violated Maldacena's consistency relation and immediately ruled out in one shot all of the simplest inflationary models. More precisely, we would have learned that one of the three standard assumptions listed above is invalid. Instead, the Planck result~\cite{Ade:2013ydc} $f_{\rm NL}^{\rm local} = 2.7\pm 5.8$ is
so far consistent with Maldacena's relation. What~\eqref{schematicintro} shows is that there are in fact an infinite number of additional checks that the simplest inflationary scenarios must pass
to be validated.

In this paper, we investigate the role of the initial state in the derivation of the consistency relations and identify the various ways in which departures from the Bunch-Davies state can result in
violations of these relations. This study is motivated in part by the recent derivation of Goldberger {\it et al.}~\cite{Goldberger:2013rsa} of Maldacena's consistency relation
as a consequence of the Ward identity from spontaneously broken dilation. Since one is primarily interested in equal-time correlation functions in cosmology, these authors
introduced a 3d Euclidean path integral over field configurations at fixed time. All of the information about the prior history is encoded in the wavefunctional. The 3d path integral approach,
and its connection to the 4d in-in path integral is reviewed in Sec.~\ref{frameworkrev}.

Surprisingly, it seems that the derivation of~\cite{Goldberger:2013rsa} applies to {\it any} gauge-invariant initial state. This seems too general, as
the authors themselves point out. Moreover, explicit calculations of the bispectrum based on particular initial states~\cite{Agarwal:2012mq}
seem to offer counterexamples. Our goal is to resolve this apparent contradiction by pinpointing the precise role of the initial state in the consistency relations. 

The derivation of the Slavnov-Taylor identity~\eqref{wardintro} is also based on the fixed-time approach~\cite{Berezhiani:2013ewa}. In particular, $\Gamma^{3{\rm d}}$ is the 3d vertex functional,
defined as usual as the Legendre transform of the connected generating functional. As argued in~\cite{Berezhiani:2013ewa} and reviewed in Sec.~\ref{STrev}, the consistency relations follow from~\eqref{wardintro} provided that the 3d vertices satisfy a certain analyticity/locality condition as $\vec{q} \rightarrow 0$. As shown in~\cite{Berezhiani:2013ewa}, this locality condition is satisfied if mode functions have constant growing-mode solutions and the initial state is the Bunch-Davies vacuum.  

More generally, to understand the role of the initial state we must relate $\Gamma^{3{\rm d}}$ back to the 4d vertex functional $\Gamma^{4{\rm d}}$ computed from the full in-in path integral. Focusing on pure initial states, for simplicity, the initial wavefunctional $\Psi \sim e^{i\mathcal{S}}$ can be thought of as an additional contribution to the action which is localized at the initial time. At tree level, we schematically write
\be
\Gamma^{4{\rm d}} = S + \mathcal{S}\,,
\ee
where $S$ is the ``bulk'' action, and $\mathcal{S}$ is the contribution from the wavefunctional. Our first result, derived in Sec.~\ref{spacediff}, is that spatial diffeomorphism invariance imposes that $S$ and $\mathcal{S}$ each satisfy a Slavnov-Taylor identity of the form~\eqref{wardintro}.

The 3d and 4d vertices are related by (see Sec.~\ref{4dto3dsubsec})
\beq
\Gamma^{3{\rm d}\,\zeta\zeta\zeta}(\vec{q},\vec{p},-\vec{q}-\vec{p}) &=&i\int_{t_0}^t {\rm d}t_1 {\rm d}t_2{\rm d}t_3 \frac{P_\zeta(q,t,t_1)}{P_\zeta(q)} \frac{P_\zeta(p,t,t_2)}{P_\zeta(p)} \frac{P_\zeta(|\vec{p}+\vec{q}|,t,t_3)}{P_\zeta(|\vec{p}+\vec{q}|)}\frac{\delta^3 S}{\delta \zeta_{\vec{q}}(t_1)\delta \zeta_{\vec{p}}(t_2)\delta \zeta_{-\vec{q}-\vec{p}}(t_3)}\nonumber\\
&+&i\frac{P_\zeta(q,t,t_0)}{P_\zeta(q)} \frac{P_\zeta(p,t,t_0)}{P_\zeta(p)} \frac{P_\zeta(|\vec{p}+\vec{q}|,t,t_0)}{P_\zeta(|\vec{p}+\vec{q}|)}C_{\vec{q},\vec{p},-\vec{q}-\vec{p}}\,,
\label{splitintro}
\eeq
where $P_\zeta(t,t_i)$ is the non-equal time 2-point function, and $C_{\vec{q},\vec{p},-\vec{q}-\vec{p}}$ is a cubic contribution from the initial wavefunctional $\mathcal{S}$.
Similarly for vertices involving tensors. This clearly displays the possible non-local contributions to $\Gamma^{3{\rm d}}$ arising from a modified initial state: 

\begin{itemize}

\item The second line is a non-Gaussian contribution from the initial state, whose $q$ dependence is not fixed by symmetries. In Sec.~\ref{NGinitial}, for example, we focus on local initial non-Gaussianities and show that the resulting contribution to $\Gamma^{3{\rm d}\,\zeta\zeta\zeta}$ is ${\cal O}(q^0)$, which violates the ${\cal O}(q^2)$ behavior required for the consistency relations. 

\item The Gaussian part of the initial state enters in the first line of~\eqref{splitintro} through non-equal-time 2-point functions. Even though $S$ vanishes as ${\cal O}(q^2)$
(assuming constant growing modes), the interplay of the modified 2-point functions and the time integral can lead to an ${\cal O}(q)$ contribution to $\Gamma^{3{\rm d}\,\zeta\zeta\zeta}$,
which violates the consistency relations. In Sec.~\eqref{gaussianinitial}, we illustrate this explicitly with Bogoliubov states.

\end{itemize}

We conclude with future directions of investigation in Sec.~\ref{conclu}.

\section{Framework}
\label{frameworkrev}

The in-in path-integral formalism, which is widely used in non-equilibrium field theory, is a powerful tool for studying the expectation values of various operators. Consider a system described by the action $S[\Phi]$ and prepared in the state with a density operator $\rho$ at some initial time $t_0$. Here, $\Phi$ collectively denotes all relevant degrees of freedom; eventually, we will take it to denote the inflaton and graviton degrees of freedom. 

\subsection{4d Formalism}
\label{4dpath}

Following the standard ``doubling'' of fields, the unified description of all correlation functions in terms of the generating functional can be obtained by introducing appropriate external currents:
\beq
Z[J^{+},J^{-}]=\int \mathcal{D}\Phi^{+}~\mathcal{D}\Phi^{-}\exp \bigg[ i\Big( S[\Phi^{+},J^{+}]-S[\Phi^{-},J^{-}] \Big) \bigg]\rho(\Phi^{+},\Phi^{-};t_0)\,,
\label{gen}
\eeq
where the time integrals run from the initial time $t_0$ to the time of evaluation $t$, the path-integral is performed over the configurations satisfying $\Phi^{+}(\vec{x},t)=\Phi^{-}(\vec{x},t)$,
and $\rho(\Phi^{+},\Phi^{-};t_0)$ represents the initial density matrix.\footnote{To be precise, $\rho(\Phi^{+},\Phi^{-};t_0)$ is the representation of the density operator in the field-eigenstate or configuration-space basis.} It is convenient to express the density matrix $\rho(\Phi^{+},\Phi^{-};t_0)$ as an exponential~\cite{Agarwal:2012mq}
\beq
\rho(\Phi^{+},\Phi^{-};t_0)\sim {\rm exp}\left( i \mathcal{S}\left[ \Phi^{+},\Phi^{-};t_0 \right] \right)\,,
\label{density}
\eeq
where $\mathcal{S}\left[ \Phi^{+},\Phi^{-};t_0 \right]^{*}=-\mathcal{S}\left[ \Phi^{-},\Phi^{+};t_0 \right]$ to ensure that $\rho$ is hermitian.
Thus $\mathcal{S}$ represents a contribution to the action localized at the initial time.

To simplify the notation we will adopt the 2-component description $\Phi^a=(\Phi^+,\Phi^-)$ and $J^a=(J^+,J^-)$, where $a=\{+,-\}$ and indices are lowered by the metric $\eta_{ab}={\rm diag}(1,-1)$. The generating functional becomes
\be
Z[J^a]=\int \mathcal{D}\Phi^a \exp\bigg[ i\Big(S[\Phi^a]+ \mathcal{S}\left[\Phi^a;t_0 \right] + \int {\rm d}^4 x J_a\Phi^a \Big)  \bigg]\,,
\label{generator}
\ee
where $S[\Phi^a] \equiv S[\Phi^+]-S[\Phi^-]$. The generator of connected diagrams, defined as usual by $W\equiv -i{\rm ln}Z$, satisfies\footnote{In~\eqref{phi}, $\Phi^a$ represents the expectation value of the field in the presence of the external current. For simplicity, the same notation as for the field itself is being used.}
\beq
\Phi^a(x)=\frac{\delta W}{\delta J_a(x)}\,.
\label{phi}
\eeq
The vertex functional will play a central role in the rest of our discussion. It is defined as usual through the Legendre transform
\beq
\Gamma^{\rm 4d}[\Phi^a]=W[J^a]-\int {\rm d}^4 x J_a \Phi^a\,,
\label{Gamma4d}
\eeq
where the `4d' label is introduced to distinguish this from the vertex functional derived from the 3d path integral to be defined shortly. 
Differentiation with respect to $\Phi$ combined with~\eqref{phi} gives the inverse relation
\beq
J_a(x) = -\frac{\delta\Gamma^{\rm 4d}}{\delta \Phi^a(x)}\,.
\label{J}
\eeq

It is straightforward to derive the Feynman rules in the usual manner. Specifically, if we differentiate~\eqref{phi} and~\eqref{J} with respect to $J$ and $\Phi$ respectively, followed by the convolution of the resulting equations, we obtain
\beq
\int {\rm d}^4z\frac{\delta^2 \Gamma^{\rm 4d}}{\delta \Phi^a(x) \delta \Phi^c(z)}\frac{\delta^2 W}{\delta J_c(z) \delta J_b(y)}=-\delta_{\,a}^b\delta^{4}(x-y)\,.
\label{4d2point}
\eeq
The Feynman rules follow from this relation by differentiating with respect to $J$ a number of times and reshuffling terms appropriately. For example, the relation between the 3-point function and the 3-point vertex is given by
\beq
\frac{\delta^3 W}{\delta J_a(x)\delta J_b(y)\delta J_c(z)}=\int {\rm d}^4x' {\rm d}^4y' {\rm d}^4z'\frac{\delta^2 W}{\delta J_a(x)\delta J_{a'}(x')}\frac{\delta^2 W}{\delta J_b(y)\delta J_{b'}(y')}\frac{\delta^2 W}{\delta J_c(z)\delta J_{c'}(z')}\nonumber \\
\times \frac{\delta^3 \Gamma^{\rm 4d}}{\delta \Phi^{a'}(x')\delta \Phi^{b'}(y')\delta \Phi^{c'}(z')}\,.
\label{4d3point}
\eeq
Such relations establish a dictionary between vertices and the connected correlators in the in-in path integral.

\subsection{3d Formalism}
\label{3dpath}

In cosmology, one is usually interested in equal-time correlation functions only. For this purpose, it is convenient to rephrase the problem in terms of a fixed-time or 3d path integral~\cite{Goldberger:2013rsa}.
In this description, the history has been integrated out, and the path integral is over field configurations at {\it fixed time} $t$. In our discussion of consistency relations, it will prove instructive to go
back and forth between the 3d and 4d descriptions.

Concretely, consider the generating function with an external source localized at some late time $t$, {\it i.e.}, with $J_a(\vec{x},t')=-i\delta_{a+}\mathcal{J}(\vec x)\delta(t'-t)$. In this case~\eqref{generator} reduces to
\beq
Z[\mathcal{J}]=\int \mathcal{D}\Phi P[\Phi,t]\exp\bigg[{\int {\rm d}^3 x\Phi(\vec{x},t)\mathcal{J}(\vec{x})}\bigg]\,.
\label{3dgenerator}
\eeq
The probability distribution $P[\Phi,t]$ is itself represented by a path integral
\beq
P[\Phi,t]= \int^{\Phi(\vec{x},t)} \mathcal{D}\Phi^a \exp\bigg[ i\Big(S[\Phi^a]+ \mathcal{S}\left[\Phi^a;t_0 \right]  \Big)  \bigg]\,,
\label{prob}
\eeq
subject to the boundary condition $\Phi^{+}(\vec{x},t)=\Phi^{-}(\vec{x},t) = \Phi(\vec{x},t)$. Although~\eqref{3dgenerator} represents a 3d path integral at fixed time $t$, its dependence on the history prior to $t$ is encoded in $P[\Phi,t]$. Following \cite{Goldberger:2013rsa}, we can define the 3d connected generating functional\footnote{The missing factor of `$i$' compared to the usual definition of $W$ is because of the Euclidean nature of the path-integral.} $\mathcal{W}\equiv {\rm ln} Z$, and the corresponding vertex functional
\beq
\Gamma^{\rm 3d}[\Phi]=\mathcal{W}-\int {\rm d}^3 x \mathcal{J}(\vec{x})\Phi(\vec{x})\,,
\label{legendre3d}
\eeq
As in the 4d case, we get the standard relations
\beq
\Phi(\vec{x})=\frac{\delta \mathcal{W}}{\delta \mathcal{J}(\vec{x})}\,; \qquad \mathcal{J}(\vec{x})=-\frac{\delta \Gamma^{\rm 3d}}{\delta \Phi(\vec{x})}\,.
\eeq
Once again, using these relations we can derive the 3d analogue of~\eqref{4d2point},
\beq
\label{2d2point}
\int {\rm d}^3z\frac{\delta^2 \Gamma^{\rm 3d}}{\delta \Phi(\vec{x}) \delta \Phi(\vec{z})}\frac{\delta^2 \mathcal{W}}{\delta \mathcal{J}(\vec{z}) \delta \mathcal{J}(\vec{y})}=-\delta^{3}(\vec{x}-\vec{y})\,,
\eeq
and similarly the analogue of~\eqref{4d3point},
\beq
\frac{\delta^3 \mathcal{W}}{\delta \mathcal{J}(\vec{x})\delta \mathcal{J}(\vec{y})\delta \mathcal{J}(\vec{z})}=\int {\rm d}^3x' {\rm d}^3y' {\rm d}^3z'\frac{\delta^2 \mathcal{W}}{\delta \mathcal{J}(\vec{x})\delta \mathcal{J}(\vec{x}')}\frac{\delta^2 \mathcal{W}}{\delta \mathcal{J}(\vec{y})\delta \mathcal{J}(\vec{y}')}\frac{\delta^2 \mathcal{W}}{\delta \mathcal{J}(\vec{z})\delta \mathcal{J}(\vec{z}')}\nonumber \\
\times \frac{\delta^3 \Gamma^{\rm 3d}}{\delta \Phi(\vec{x}')\delta \Phi(\vec{y}')\delta \Phi(\vec{z}')}\,.
\label{3d3point}
\eeq
In the above, $\delta^n \mathcal{W}/\delta\Phi^n$ represents the equal-time connected $n$-point function, and all fields are evaluated at time $t$. 

Since the 3d and 4d formulations must describe the same physics, the path integrals~\eqref{generator} and~\eqref{3dgenerator} generate 
identical equal-time correlation functions. This implies a relation between the vertex functions $\Gamma^{\rm 3d}$ and $\Gamma^{\rm 4d}$ which will be useful
for the forthcoming discussion. This relation is most elegantly represented in Fourier space. Our starting point is the equal time 3-point correlator
derived in the 4d formalism, which follows from~\eqref{4d3point}. In Fourier space, it is given by
\beq
\la \Phi_{\vec{q}}(t) \Phi_{\vec{p}}(t) \Phi_{-\vec{q}-\vec{p}}(t) \ra '= i\int {\rm d}t_1 {\rm d}t_2{\rm d}t_3 G_{+a}(q,t,t_1) G_{+b}(p,t,t_2)G_{+c}(|\vec{p}+\vec{q}|,t,t_3)\nonumber \\
\times\frac{\delta^3 \Gamma^{4{\rm d}}}{\delta \Phi^a_{\vec{q}}(t_1)\delta \Phi^{b}_{\vec{p}}(t_2)\delta \Phi^c_{-\vec{q}-\vec{p}}(t_3)}\,,
\label{4d3p}
\eeq
where $G_{+a}(q,t,t_i)$ denotes the non-equal-time 2-point function connecting $\Phi^a(t_i)$ to $\Phi^+(t) \equiv \Phi(t) $. Moreover, as usual $\langle \ldots \rangle'$ is an on-shell correlation with the delta function removed.\footnote{The precise statement is $\langle {\cal O}_{\vec{k}_1,\ldots,\vec{k}_N}\rangle = (2\pi)^3 \delta^3(\vec{k}_1 + \ldots + \vec{k}_N)\langle {\cal O}_{\vec{k}_1,\ldots,\vec{k}_N}\rangle'$.} In the 3d Euclidean approach, meanwhile, the analogous result follows from~\eqref{3d3point},
\beq
\la \Phi_{\vec{q}}(t) \Phi_{\vec{p}}(t) \Phi_{-\vec{q}-\vec{p}}(t) \ra '=G(q,t)G(p,t)G(|\vec{q}+\vec{p}|,t)\frac{\delta^3 \Gamma^{3{\rm d}}}{\delta \Phi_{\vec{q}}~\delta \Phi_{\vec{p}}~\delta \Phi_{-\vec{q}-\vec{p}}}\,,
\label{3d3p}
\eeq
where $G(q,t)$ represents the equal-time 2-point function. Equating~\eqref{4d3p} and~\eqref{3d3p}, we obtain a relation among 3d and 4d vertices:
\ba
\nonumber
& & \frac{\delta^3 \Gamma^{3{\rm d}}}{\delta \Phi_{\vec{q}}~\delta \Phi_{\vec{p}}~\delta \Phi_{-\vec{q}-\vec{p}}}= i\int {\rm d}t_1 {\rm d}t_2{\rm d}t_3 \frac{G_{+a}(q,t,t_1)}{G(q,t)} \frac{G_{+b}(p,t,t_2)}{G(p,t)} \frac{G_{+c}(|\vec{p}+\vec{q}|,t,t_3)}{G(|\vec{p}+\vec{q}|,t)}\\
& & ~~~~~~~~~~~~~~~~~~~~~~~~~~~~~~~~~~~~~~~~~~~~~~~~~~~~~~~~~~~~~~~~~~~\times\frac{\delta^3 \Gamma^{4{\rm d}}}{\delta \Phi^a_{\vec{q}}(t_1)\delta \Phi^{b}_{\vec{p}}(t_2)\delta \Phi^c_{-\vec{q}-\vec{p}}(t_3)}\,.~
\label{convert1}
\ea
This relation will play a key role in understanding the possible violations of the consistency condition. 

In the following sections we will concentrate on pure initial states for concreteness. In this case, the density matrix factorizes as
\beq
\rho(\Phi^{a};t_0)\sim \exp\bigg[i \Big(\mathcal{S}\left[ \Phi^{+};t_0 \right]-\mathcal{S}\left[\Phi^{-};t_0 \right] \Big)\bigg]\,.
\label{rhopure}
\eeq
As a result, $G_{ab}$ becomes diagonal, and~\eqref{convert1} reduces to
\beq
\frac{\delta^3 \Gamma^{3{\rm d}}}{\delta \Phi_{\vec{q}}~\delta \Phi_{\vec{p}}~\delta \Phi_{-\vec{q}-\vec{p}}}= i\int {\rm d}t_1{\rm d}t_2{\rm d}t_3 \frac{G_{++}(q,t,t_1)}{G(q,t)} \frac{G_{++}(p,t,t_2)}{G(p,t)} \frac{G_{++}(|\vec{p}+\vec{q}|,t,t_3)}{G(|\vec{p}+\vec{q}|,t)}\nonumber \\
\times\frac{\delta^3 \Gamma^{4{\rm d}}}{\delta \Phi^+_{\vec{q}}(t_1)\delta \Phi^{+}_{\vec{p}}(t_2)\delta \Phi^+_{-\vec{q}-\vec{p}}(t_3)}\,.~
\label{convert2}
\eeq
When applying~\eqref{convert2} to the cosmological case, $\Phi$ will be replaced by the appropriate gravitational degrees of freedom, namely scalar/tensor perturbations 
on an FRW background. In that case, $G(q,t)$ will represent the scalar/tensor power spectrum, while $G_{++}(q,t,t_i)$ will become a scalar/tensor 2-point Green's function.

\section{Consistency Conditions from Slavnov-Taylor Identity}
\label{STrev}

In this Section, we briefly review the derivation of~\cite{Berezhiani:2013ewa} of the consistency relations based on the Slavnov-Taylor identity
for spatial reparametrization invariance. The derivation applies to any spatially-flat homogeneous background generated by a single scalar field:
\beq
\text{d}s^2=-\text{d}t^2+a^2(t)\text{d}\vec{x}^2\,; \qquad \phi=\bar{\phi}(t)\,.
\eeq
Following~\cite{Berezhiani:2013ewa}, we work in uniform-density gauge, where the scalar field is unperturbed, 
\beq
\delta\phi\equiv \phi(\vec{x},t)-\bar{\phi}(t)=0\,,
\eeq
and the spatial metric is 
\be
H_{ij} = a^2(t)e^{2\zeta(\vec{x},t)} \left(e^{\gamma(\vec{x},t)}\right)_{ij}\,,
\ee
where $\gamma^i_{\,i} = 0$, $\partial^i\gamma_{ij} = 0$. Scalar modes are encoded in the curvature perturbation $\zeta$, while tensor modes (gravitational waves) are encoded in the transverse, traceless
perturbation $\gamma_{ij}$. This gauge choice represents unitary gauge from the point of view of time reparametrization invariance, hence spatial diffeomorphisms are in fact the only symmetries to consider.

The Slavnov-Taylor identity is most elegantly derived in the 3d fixed-time path integral of Sec.~\ref{3dpath}. We imagine that the auxiliary lapse function and shift vector have been integrated out using the constraints, hence the remaining degrees of freedom are the scalar and tensor perturbations of the spatial metric. The fixed-time path integral is then of the form
\be
Z[\mathcal{J}_\zeta,\mathcal{J}_\gamma] = \int \mathcal{D}\zeta \,\mathcal{D}\gamma_{ij} \mathcal{P}[\zeta,\gamma,t] \exp\bigg[ \int {\rm d}^3 x\left(\zeta \mathcal{J}_\zeta +  \gamma_{ij} \mathcal{J}^{ij}_\gamma\right) \bigg]\,,
\label{3d}
\ee
where the $\mathcal{J}$'s represent scalar and tensor external currents. The probability distribution $\mathcal{P}[\zeta,\gamma,t]$ encodes the information about the action describing the theory, as well as the initial state in which quantum fluctuations are created. This state is usually considered to be the Bunch-Davies vacuum. Instead, we consider a completely general initial state specified at some initial time $t_0$. Our only restriction is that {\it the initial state respect local diffeomorphism invariance}. By ``local'' diffeomorphisms, we mean diffeomorphisms that fall off suitably fast at spatial infinity.\footnote{In particular, we do not impose that the initial state be invariant under {\it global} diffeormorphisms, which would be a much stronger requirement.}

To lowest order in the tensors, the spatial reparametrization-invariance of $Z[\mathcal{J}_\zeta,\mathcal{J}_\gamma]$ leads to a variational differential equation for the 3d vertex functional~\cite{Berezhiani:2013ewa}:
\beq
2\p_j \left( \frac{1}{6} \delta_{ij}\frac{\delta\Gamma^{3{\rm d}}}{\delta\zeta}+\frac{\delta \Gamma^{3{\rm d}}}{\delta \gamma_{ij}} \right)=\p_i \zeta \frac{\delta \Gamma^{3{\rm d}}}{\delta \zeta} + \ldots
\label{ward}
\eeq
The ellipses include the contribution from the gauge-fixing term, and terms that are higher order in the tensors. At tree level, these terms can be ignored for the remainder of our discussion. See~\cite{Berezhiani:2013ewa} for details. Although~\eqref{ward} was originally derived for the Bunch-Davies initial state, it holds more generally for any initial state that is invariant under local spatial diffeomorphisms.

As in~\cite{Berezhiani:2013ewa}, we focus on consistency relations relating 3-point functions with a soft mode to 2-point functions without the soft mode. Moreover, the hard-momentum modes
are assumed to be scalars. We must therefore perform the functional differentiation of~\eqref{ward} with respect to $\zeta(\vec{x}_1,t)$ and $\zeta(\vec{x}_2,t)$, and afterwards set all fields to zero. The result
in Fourier space is
\beq
q^j\left(\frac{1}{3}\delta_{ij}\Gamma^{3{\rm d}\,\zeta\zeta\zeta}(\vec{q},\vec{p},-\vec{q}-\vec{p})+2 \Gamma_{ij}^{3{\rm d}\,\gamma\zeta\zeta}(\vec{q},\vec{p},-\vec{q}-\vec{p})\right)=q_i \Gamma_\zeta^{3{\rm d}}( p )-p_i\bigg( \Gamma_\zeta^{3{\rm d}}(|\vec{q}+\vec{p}|)-\Gamma_\zeta^{3{\rm d}}( p ) \bigg)\,,
\label{wardf}
\eeq
where $\Gamma^{3{\rm d}\,\zeta\zeta\zeta}$ is the vertex with 3 scalars, $\Gamma_{ij}^{3{\rm d}\,\gamma\zeta\zeta}$ is the vertex with 2 scalars and 1 tensor, and $\Gamma_\zeta^{3{\rm d}}$ is the inverse scalar propagator.\footnote{More precisely, the relations are
\beq
\nonumber
\int {\rm d}^3x_1\, {\rm d}^3x_2 e^{-i(\vec{p}_1\cdot\vec{x}_1 + \vec{p}_2\cdot \vec{x}_2)}\left.\frac{\delta^2 \Gamma^{3{\rm d}}}{\delta \zeta(\vec{x}_1)\delta \zeta(\vec{x}_2)}\right\vert_{\zeta = \gamma =0} &=& (2\pi)^3 \delta^3 (\vec{p}_1 + \vec{p}_2)\Gamma_\zeta^{3{\rm d}}( p_1 )\,;\\
\nonumber
\int {\rm d}^3 x_1 \, {\rm d}^3 x_2\, {\rm d}^3 x_3 e^{-i\sum \vec{p}_i\cdot \vec{x}_i}\left.\frac{\delta^3\Gamma^{3{\rm d}}}{\delta \zeta(\vec{x}_1) \delta \zeta(\vec{x}_2) \delta \zeta(\vec{x}_3)}\right\vert_{\zeta = \gamma =0} &=& (2\pi)^3\delta^3(\vec{p}_1+\vec{p}_2 + \vec{p}_3)\Gamma^{3{\rm d}\,\zeta\zeta\zeta}(\vec{p}_1,\vec{p}_2,\vec{p}_3)\,;\\
\int {\rm d}^3 x_1 \, {\rm d}^3 x_2\, {\rm d}^3 x_3 e^{-i\sum \vec{p}_i\cdot \vec{x}_i} \left.\frac{\delta^3\Gamma^{3{\rm d}}}{\delta \gamma_{ij}(\vec{x}_1) \delta \zeta(\vec{x}_2) \delta \zeta(\vec{x}_3)}\right\vert_{\zeta = \gamma =0}  &=& (2\pi)^3\delta^3(\vec{p}_1+\vec{p}_2 + \vec{p}_3)\Gamma_{ij}^{3{\rm d}\,\gamma\zeta\zeta}(\vec{p}_1,\vec{p}_2,\vec{p}_3)\,.~~~~~
\eeq
}
This relation constrains the 3-point vertices, which are related to 3-point correlation functions as follows
\beq
\nonumber
\langle \zeta_{\vec{q}} \zeta_{\vec{p}} \zeta_{-\vec{q} - \vec{p}} \rangle' &=&P_\zeta(q) P_\zeta( p )P_\zeta(|\vec{q} + \vec{p}|) \Gamma^{3{\rm d}\,\zeta\zeta\zeta}(\vec{q},\vec{p},-\vec{q} - \vec{p}) \,;\\
\langle \gamma^{ij}_{\vec{q}}\zeta_{\vec{p}} \zeta_{-\vec{q} - \vec{p}} \rangle' &=& \hat{P}^{ijk\ell}(\hat{q})P_\gamma(q) P_\zeta( p )P_\zeta(|\vec{q} + \vec{p}|)\Gamma_{k\ell}^{3{\rm d}\,\gamma\zeta\zeta}(\vec{q},\vec{p},-\vec{q} - \vec{p})\,,
\label{correl}
\eeq
where $\hat{P}_{ijk\ell}= P_{ik} P_{j\ell} + P_{i\ell} P_{jk} - P_{ij}P_{k\ell}$ is the transverse, traceless tensor appearing in the graviton propagator, and
$P_{ij} = \delta_{ij} - \hat{q}_i\hat{q}_j$ is the transverse projector. 

The most general solution to~\eqref{wardf} can be written as
\beq
P_\zeta(p )P_\zeta(|\vec{q}+\vec{p}|)\left(\frac{1}{3}\delta_{ij}\Gamma^{3{\rm d}\,\zeta\zeta\zeta}(\vec{q},\vec{p},-\vec{q}-\vec{p})+2\Gamma_{ij}^{3{\rm d}\,\gamma\zeta\zeta}(\vec{q},\vec{p},-\vec{q}-\vec{p})\right)= K_{ij}(\vec{p},\vec{q}) + A_{ij}(\vec{p},\vec{q}) \,,
\label{soln}
\eeq
where 
\beq
K_{ij} \equiv -\delta_{ij} P_\zeta( p ) -p_{(i}\frac{\p P_\zeta(p)}{\p p^{j)}}-\sum_{n=1}^{\infty} \frac{q_{\alpha_1}\ldots q_{\alpha_n}}{n!}\left[ \delta_{ij}\frac{\p ^n}{\p p_{\alpha_1}\ldots \p p_{\alpha_n}}+\frac{p_i}{n+1}\frac{\p ^{n+1}}{\p p_j \p p_{\alpha_1}\ldots \p p_{\alpha_n}}\right. \nonumber \\
\left.  +\frac{p_j}{n+1}\frac{\p ^{n+1}}{\p p_i \p p_{\alpha_1}\ldots \p p_{\alpha_n}} -\frac{p_{\alpha_1}}{n+1}\frac{\p ^{n+1}}{\p p_i \p p_j \p p_{\alpha_2}\ldots \p p_{\alpha_n}}\right]P_\zeta(p) \,,
\label{Kdef}
\eeq
and $A_{ij}$ is an arbitrary symmetric and transverse matrix:
\be
q^jA_{ij}(\vec{p},\vec{q})=0\,.
\ee
Note that $K_{ij}(\vec{p},\vec{q})$ is regular in the $q\to 0$ limit and encodes the model-independent, analytic part of the vertex functionals. The model-dependent part, which may be non-analytic, enters through 
the arbitrary symmetric and transverse array $A_{ij}(\vec{p},\vec{q})$. The most general $A_{ij}$ with these properties built out of $\vec{p}$, $\vec{q}$, Kronecker $\delta$'s and Levi-Cevita symbols is~\cite{Berezhiani:2013ewa}
\beq
A_{ij}=\epsilon _{ikm}\epsilon_{j\ell n}q^k q^\ell\bigg(a(\vec{p},\vec{q}) \delta^{mn}+b(\vec{p},\vec{q})p^mp^n\bigg)\,,
\label{ambig}
\eeq
where $a$ and $b$ are arbitrary scalar functions of the momenta. 

So far the derivation is completely general. The only assumption that went into~\eqref{soln} is the spatial reparametrization invariance of the probability function $\mathcal{P}[\zeta,\gamma,t]$,
which encodes information about the action and the initial densitity matrix. In particular,~\eqref{soln} holds irrespective of whether $\zeta,\gamma$ have constant growing modes
outside the horizon, or whether the initial state is highly non-Gaussian. In order to translate~\eqref{soln} into consistency relations for correlation functions, we must make a technical
assumption about the analytic behavior of $A_{ij}$ in the $\vec q\to 0$ limit. The key assumption is that the functions $a$ and $b$ are both analytic in $q$, such that $A_{ij}$ starts at order $q^2$:
\be
A_{ij} = {\cal O}(q^2)\,.
\label{localityassumption}
\ee
In this case, when converting~\eqref{soln} to correlation functions via~\eqref{correl}, one can project out $A_{ij}$ order by order in $q$ to obtain consistency relations schematically of the form
\be
\lim_{\vec{q}\rightarrow 0} {\partial^n \over \partial q^n}
\left(  \frac{\langle \zeta_{\vec q}\zeta_{\vec{p}}\zeta_{-\vec{q}-\vec{p}} \rangle'}{P_\zeta (q)} + \frac{\langle \gamma_{\vec q} \zeta_{\vec{p}}\zeta_{-\vec{q}-\vec{p}} \rangle'}{P_\gamma (q)} \right)
\sim - {\partial^n \over \partial p^n}P_\zeta (p)  \, .
\label{schematic}
\ee
See~\cite{Hinterbichler:2013dpa,Berezhiani:2013ewa} for the explicit form of these identities.

In~\cite{Berezhiani:2013ewa}, it was argued that the locality assumption~\eqref{localityassumption} is satisfied if the initial state is the Bunch-Davies state and the field perturbations
have constant growing-mode solutions in the $\vec{q}\rightarrow 0$ limit. Physically, this can be understood by going back to the 4d action $S[\zeta,\gamma]$, which at tree level coincides with the 4d
vertex functional $\Gamma^{4{\rm d}}[\zeta,\gamma]$. Assuming that the original gravity + scalar field action is itself local, then the only non-localities in $\Gamma^{4{\rm d}}$ arise from integrating out the lapse function and shift vector via the constraints. As long as the fields have constant growing-mode solutions, $\Gamma^{4{\rm d}}$ becomes local in $q$. However, this is not sufficient,
since~\eqref{localityassumption} is a locality requirement on the 3d vertex functional, instead of the 4d one. The translation, given in~\eqref{convert1}, involves equal-time and non-equal-time
Green's functions. We argued in~\cite{Berezhiani:2013ewa} that, in the case of the Bunch-Davies state, an analytic $\Gamma^{4{\rm d}}$ does translate to an analytic $\Gamma^{3{\rm d}}$,
such that~\eqref{localityassumption} is satisfied and the consistency conditions hold. Our goal in the next Section is to understand how this story changes once we allow general initial states.

\section{Ways to Violate the Consistency Relations}
\label{ways}

In this Section, we dissect the possible ways in which~\eqref{localityassumption} can fail to hold, and the consistency relations can be violated, for general initial states.
Although the Slavnov-Taylor identity is most simply derived in the 3d framework, as reviewed in Sec.~\ref{STrev}, to get further insights we must trace back the possible
consistency violations to properties of the 4d vertex functional. 

For concreteness, we focus here on pure initial states, though most of our results straightforwardly generalize to the mixed case. 
In the pure case, the relation between 3-point interaction vertices in the 3d and 4d pictures is given by~\eqref{convert2}.
Applying this relation to the cubic scalar vertex, we obtain
\beq
\Gamma^{3{\rm d}\,\zeta\zeta\zeta}(\vec{q},\vec{p},-\vec{q}-\vec{p}) = i\int_{t_0}^t {\rm d}t_1 {\rm d}t_2{\rm d}t_3 \frac{P_\zeta(q,t,t_1)}{P_\zeta(q)} \frac{P_\zeta(p,t,t_2)}{P_\zeta(p)} \frac{P_\zeta(|\vec{p}+\vec{q}|,t,t_3)}{P_\zeta(|\vec{p}+\vec{q}|)}\nonumber\\
\times\frac{\delta^3 \Gamma^{4{\rm d}}}{\delta \zeta_{\vec{q}}(t_1)\delta \zeta_{\vec{p}}(t_2)\delta \zeta_{-\vec{q}-\vec{p}}(t_3)}\,,
\label{convert}
\eeq
where $P_\zeta(p) \equiv P_\zeta(p,t)$ is the scalar power spectrum, and $P_\zeta(t,t_i)$ is the non-equal time 2-point function.
Similarly for vertices involving tensors. Notice that momentum conserving delta functions have been removed, hence these are on-shell interaction vertices.
As the discussion at the end of Sec.~\ref{STrev} already suggests, we will find two possible sources of violations:

\begin{itemize}

\item The initial state may itself encode large non-Gaussianities. This will give a non-analytic contribution to $\Gamma^{4{\rm d}}$, which will translate to a non-analytic contribution to $\Gamma^{3{\rm d}}$ via~\eqref{convert}. We will study this possibility in Sec.~\ref{NGinitial}.

\item Even if the initial state is Gaussian, it may modify the non-equal-time power spectra $P_\zeta(t,t_i)$ appearing in~\eqref{convert}. This in turn can spoil the link between analyticity of $\Gamma^{4{\rm d}}$ and analyticity of $\Gamma^{3{\rm d}}$. In particular, in Sec.~\ref{gaussianinitial} we will see with explicit examples of Bogoliubov states that an analytic $\Gamma^{4{\rm d}}$ can map to a non-analytic $\Gamma^{3{\rm d}}$, thereby violating~\eqref{localityassumption}. 

\end{itemize}

\subsection{Spatial Diffeomorphism Invariance}
\label{spacediff}

We begin by studying the implications of spatial diffeomorphism invariance for $\Gamma^{4{\rm d}}$. 
Recall that in the ADM formalism, the metric components are decomposed into lapse function $N$, shift vector $N^j$ and spatial metric $H_{ij} = a^2(t)\left(\delta_{ij} + h_{ij}\right)$.
In uniform-density gauge, the metric perturbations are further decomposed into scalar and tensor modes via $h_{ij} \equiv e^{2\zeta}\left(e^{\gamma}\right)_{ij} - \delta_{ij}$,
however for convenience let us stick to $h_{ij}$ for a while.

For pure initial states, the density matrix factorizes and is given by~\eqref{rhopure}.
At tree level, we can decompose $\Gamma^{4{\rm d}}$ as follows:
\beq
\Gamma^{\rm 4d}[h^+,h^-]=S[h^+]-S[h^-] + \mathcal{S}[h^+;t_0] - \mathcal{S}[h^-;t_0] \,.
\label{dissect}
\eeq
Here, $S$ is the diffeomorphism-invariant action of the theory. (Technically, it is reparametrization-invariant up to a gauge-fixing term, however
the gauge-fixing term can be ignored at tree level~\cite{Berezhiani:2013ewa}.) Meanwhile, $\mathcal{S}$ is the contribution
from the density matrix, as defined in~\eqref{rhopure}. Note that in writing down~\eqref{dissect} we have ignored the possible dependence
on $N$ and $N_j$. For the ``bulk'' action $S$, we imagine that these auxiliary fields have been integrated out using the constraints. For the initial
state contribution $\mathcal{S}$, we recall that quantum-mechanically the primary constraints of General Relativity reduce to~\cite{DeWitt:1967yk}
\beq
\Pi_N |\Psi\ra=0\,;\qquad \Pi_{N_j}|\Psi\ra=0\,,
\label{GRcons}
\eeq
where $\Pi_N$ and $\Pi_{N _j}$ are the conjugate momenta to $N$ and $N_j$, respectively. 
Equations~\eqref{GRcons} state that the shift and lapse are non-dynamical degrees of freedom. In the
configuration basis, they imply that the wavefunctional is independent of $N$ and $N_j$.\footnote{Similarly, physically-allowed density operators must be constructed from $|\Psi\ra$'s
satisfying~\eqref{GRcons}, and the corresponding density matrix are therefore independent of $N$ and $N_j$. To conclude,
$\mathcal{S}$ only depends on $h^\pm$ evaluated at the initial time $t_0$.} This makes physical sense.
Since all dynamical degrees of freedom are described by $h_{ij}$, while $N$ and $N_j$ are merely auxiliary fields. States constructed
out of the vacuum state by applying creation operators of the physical particles would contain $h_{ij}$ only. Hence $|\Psi\rangle$
must be independent of $N$ and $N_j$. 

In order to understand the possible impact of the modified initial state on the validity of consistency relations, it is imperative to
understand the properties of $\mathcal{S}$ imposed by the consistency of the theory. The momentum constraints embody the requirement of spatial diffeomorphism invariance on the possible initial states \cite{DeWitt:1967yk,Pimentel:2013gza}. Demanding that $\mathcal{S}$ be invariant
under those symmetries amounts to the requirement
\beq
\int {\rm d}^3 x \delta h^{\pm}_{ij} (\vec{x},t_0)\frac{\delta \mathcal{S}[h^\pm;t_0]}{\delta h^{\pm}_{jk}(\vec{x},t_0)} = 0\,,
\label{initialward0}
\eeq
where $\delta h^{\pm}_{ij}$ is the transformation under diffeomorphisms:
\beq
\delta h^{\pm}_{ij}=\p_i \xi_j+\p_j \xi_i+\xi^k\p_k h^{\pm}_{ij}+h^{\pm}_{ik}\p_j \xi^k+h^{\pm}_{jk}\p_i \xi^k\,.
\label{sym}
\eeq
Substituting this explicit form of the transformation and using the fact that $\xi^i$ is arbitrary,~\eqref{initialward0} implies\footnote{Allowing for mixed states, the more general requirement is
\beq
\sum_{\pm}\left[2\p_j\frac{\delta \mathcal{S}[h^+,h^-;t_0]}{\delta h^{\pm}_{jk}(\vec{x},t_0)}-\p_k h^{\pm}_{ij}(\vec{x},t_0) \frac{\delta \mathcal{S}[h^+,h^-;t_0]}{\delta h^{\pm}_{ij}(\vec{x},t_0)}+2\p_j \left( h^{\pm}_{ik}(\vec{x},t_0)\frac{\delta \mathcal{S}[h^+,h^-;t_0]}{\delta h^{\pm}_{jk}(\vec{x},t_0)} \right)\right]=0\,.
\label{initialward}
\eeq
}
\beq
2\p_j\frac{\delta \mathcal{S}[h^\pm;t_0]}{\delta h^{\pm}_{jk}(\vec{x},t_0)}-\p_k h^{\pm}_{ij}(\vec{x},t_0) \frac{\delta \mathcal{S}[h^\pm;t_0]}{\delta h^{\pm}_{ij}(\vec{x},t_0)}+2\p_j \left( h^{\pm}_{ik}(\vec{x},t_0)\frac{\delta \mathcal{S}[h^\pm;t_0]}{\delta h^{\pm}_{jk}(\vec{x},t_0)} \right) =0\,.
\label{initialwardpure}
\eeq
By varying this a number of times with respect to $h$, we obtain all possible constraints on the vertices coming from diffeomorphism invariance. Perhaps not surprisingly,
the resulting identities are identical to Slavnov-Taylor identities derived in Sec.~\ref{STrev}. In other words, the initial state must itself satisfy the consistency relations! The severity of this relation depends on the assumed analyticity properties of the wavefunctional, in complete analogy with the discussion of the previous section. In particular, in case of the analytic $\mathcal{S}$ the Gaussian and non-Gaussian parts of the initial state/density matrix are interwoven:  the squeezed limit of the $n^{\rm th}$-order term of $\mathcal{S}$ is related to
$(n-1)^{\rm th}$-order term. However, in case of the more general (\textit{i.e.} non-analytic) states, the constraint \eqref{initialwardpure} becomes rather mild. Taking all this into account, the simplest way to construct an initial wavefunction consistent with the quantum constraints is
to write $\mathcal{S}$ as a general functional of curvature scalars built out of $h^{\pm}_{jk}$. 

Similarly, the ``bulk'' action $S[h^\pm]$ satisfies an identical identity:\footnote{Recall that we have ignored the gauge-fixing term in our tree-level considerations.
More generally,~\eqref{action} would be satisfied by the action minus the gauge-fixing term.}
\beq
2\p_j\frac{\delta S[h^\pm]}{\delta h_{jk}^\pm(x)}-\p_kh_{ij}^\pm(x) \frac{\delta S[h^\pm]}{\delta h_{ij}^\pm(x)}+2\p_j \left( h_{ik}^\pm(x)\frac{\delta S[h^\pm]}{\delta h_{jk}^\pm(x)} \right)=0\,.
\label{action}
\eeq
And since $\Gamma^{\rm 4d}$ is just a linear combination of $S$ and $\mathcal{S}$, it also satisfies this identity. 
The rest of the analysis proceeds as in Sec.~\ref{STrev}. Namely, upon decomposing $h_{ij}$ into $\zeta$ and $\gamma_{ij}$,
we find that $\Gamma^{\rm 4d}$ satisfies an equation analogous to~\eqref{ward}:
\beq
2\p_j \left( \frac{1}{6} \delta_{ij}\frac{\delta\Gamma^{4{\rm d}}}{\delta\zeta^\pm}+\frac{\delta \Gamma^{4{\rm d}}}{\delta \gamma_{ij}^\pm} \right)=\p_i \zeta^\pm \frac{\delta \Gamma^{4{\rm d}}}{\delta \zeta^\pm} + \ldots
\label{ward4d}
\eeq
where the ellipses again encode irrelevant terms. This can be varied twice with respect to $\zeta$ to obtain a Slavnov-Taylor identity of the form~\eqref{wardf}
for the cubic part of $\Gamma^{\rm 4d}$. For instance, the cubic vertex for scalars receives contributions from both $S$ and $\mathcal{S}$, 
\beq
\frac{\delta^3 \Gamma^{4{\rm d}}}{\delta \zeta_{\vec{q}}(t_1)\delta \zeta_{\vec{p}}(t_2)\delta \zeta_{-\vec{q}-\vec{p}}(t_3)}=\frac{\delta^3 S}{\delta \zeta_{\vec{q}}(t_1)\delta \zeta_{\vec{p}}(t_2)\delta \zeta_{-\vec{q}-\vec{p}}(t_3)}+\frac{\delta^3 \mathcal{S}}{\delta \zeta_{\vec{q}}(t_1)\delta \zeta_{\vec{p}}(t_2)\delta \zeta_{-\vec{q}-\vec{p}}(t_3)}\,,
\label{ing}
\eeq
and similarly for vertices involving tensors. The validity of the consistency relations has now been traced back to the analyticity properties of $\Gamma^{\rm 4d}$, and hence
of $S$ and $\mathcal{S}$, in the $\vec{q}\rightarrow 0$ limit. 

\subsection{From 4d to 3d}
\label{4dto3dsubsec}

If perturbations have constant growing mode solutions as $\vec{q}\rightarrow 0$, then the bulk action $S$ will become local in this limit~\cite{Berezhiani:2013ewa}. For instance, non-local contributions to $S$ at cubic order arise from integrating out the shift vector, whose solution (at linear order) includes
\be
N_i  \supset -a^2\frac{\dot{H}}{H^2}  \frac{q_i}{q^2}\dot{\zeta}\,.
\ee
For adiabatic modes, however, $\dot{\zeta}\propto q^2$, and this contribution becomes local as $\vec{q}\rightarrow 0$. Conversely, in models where $\zeta$ is not constant outside the horizon ({\it e.g.}, because of background instabilities~\cite{Khoury:2008wj}), the action remains non-local even as $\vec{q}\rightarrow 0$. In other words, adiabaticity implies that the cubic part of $S$ is fixed up to $q^2$ order by symmetries.

However, a similar argument cannot be made for $\delta^3 \mathcal{S}/\delta \zeta^3$. The late-time behavior of mode functions clearly has no relevance
for this contribution which is localized at the initial time. In general, therefore, the cubic vertex of $\mathcal{S}$ is model-dependent to all orders in $q$. 
In particular, it may contain large initial non-Gaussianities, which should invalidate the consistency relations. 

To make these statements more concrete, consider the scalar cubic part of a generic initial state, which we assume is translation invariant:
\ba
\nonumber
& & \mathcal{S}_3=\frac{1}{3!}\int {\rm d}^3k_1{\rm d}^3k_2 \bigg( C_{\vec{k}_1,\vec{k}_2,-\vec{k}_1-\vec{k}_2} \zeta_{\vec k_1}^{+}(t_0)\zeta_{\vec k_2}^{+}(t_0)\zeta_{-\vec{k}_1-\vec{k}_2}^{+}(t_0)  \\
& & ~~~~~~~~~~~~~~~~~~~~~~~~~~~~~~~ -C^*_{\vec{k}_1,\vec{k}_2,-\vec{k}_1-\vec{k}_2} \zeta_{\vec k_1}^{-}(t_0)\zeta_{\vec k_2}^{-}(t_0)\zeta_{-\vec{k}_1-\vec{k}_2}^{-}(t_0)\bigg)\,.
\label{cubic}
\ea
This results into the following on-shell expression,
\beq
\frac{\delta^3 \mathcal{S}}{\delta \zeta_{\vec{q}}(t_1)\delta \zeta_{\vec{p}}(t_2)\delta \zeta_{-\vec{q}-\vec{p}}(t_3)}=\delta(t_1-t_0)\delta(t_2-t_0)\delta(t_4-t_0)C_{\vec{q},\vec{p},-\vec{q}-\vec{p}}\,.
\eeq
Putting this together with \eqref{ing} and \eqref{convert}, we obtain the explicit contribution to the cubic part of the 3d vertex functional:
\beq
\Gamma^{3{\rm d}\,\zeta\zeta\zeta}(\vec{q},\vec{p},-\vec{q}-\vec{p}) &=&i\int_{t_0}^t {\rm d}t_1 {\rm d}t_2{\rm d}t_3 \frac{P_\zeta(q,t,t_1)}{P_\zeta(q)} \frac{P_\zeta(p,t,t_2)}{P_\zeta(p)} \frac{P_\zeta(|\vec{p}+\vec{q}|,t,t_3)}{P_\zeta(|\vec{p}+\vec{q}|)}\frac{\delta^3 S}{\delta \zeta_{\vec{q}}(t_1)\delta \zeta_{\vec{p}}(t_2)\delta \zeta_{-\vec{q}-\vec{p}}(t_3)}\nonumber\\
&+&i\frac{P_\zeta(q,t,t_0)}{P_\zeta(q)} \frac{P_\zeta(p,t,t_0)}{P_\zeta(p)} \frac{P_\zeta(|\vec{p}+\vec{q}|,t,t_0)}{P_\zeta(|\vec{p}+\vec{q}|)}C_{\vec{q},\vec{p},-\vec{q}-\vec{p}}\,.
\label{split}
\eeq
This expression clearly displays the possible non-analytic contributions to $\Gamma^{3{\rm d}}$ arising from modifications to the initial state.
The second line represents the non-Gaussian contribution from the initial state, whose $q$ dependence is not fixed by symmetries as we
have seen. The initial state also plays a role in the first line, through the non-equal time 2-point functions. Even if the cubic part of $S$ is
local as $\vec{q}\rightarrow 0$, the time integral of the non-equal time 2-point functions can result in non-analytic contributions to
$\Gamma^{3{\rm d}}$. In the remainder of the Section, we will study explicit examples of these two sources of non-analyticity.

\subsection{Initial Local Non-Gaussianity}
\label{NGinitial}

If the initial state includes non-Gaussianities, then clearly we do not expect the consistency relations to hold. Usually, one has in mind
that the initial state is specified at the onset of inflation. In this context, the initial state (and the extent to which it is non-Gaussian) parametrizes our ignorance
about initial conditions for inflation. Alternatively, nothing prevents us from setting the initial time to be much later in the evolution.
A classic example is a multi-field scenario, such as the curvaton~\cite{Lyth:2001nq} or variable-decay~\cite{Kofman:2003nx,Dvali:2003em} mechanisms,
where the initial time might naturally coincide with the onset of adiabatic evolution, {\it i.e.}, once the additional degrees of freedom have all decayed away.
In this case, the non-Gaussianities generated by the additional fields can only appear in the initial state.

For concreteness, let us consider {\it local} initial non-Gaussianity, which is the shape of interest for multi-field scenarios. 
By definition, the 3-point amplitude is set by the well-known (constant and model-dependent) $f_{\rm NL}$ parameter:
\beq
\langle \zeta_{\vec{q}} \zeta_{\vec{p}} \zeta_{-\vec{p}-\vec{q}} \rangle ' =  \frac{6}{5} f_{\rm NL}\bigg( P_\zeta (q) P_\zeta ( p) 
+ P_\zeta (q) P_\zeta (|\vec{p} + \vec{q}|) + P_\zeta ( p) P_\zeta (|\vec{p} + \vec{q}|)\bigg) \,.
\label{curvaton}
\eeq
Using~\eqref{correl}, we can infer the contribution to the 3d vertex functional. In the squeezed limit, it is given by
\be
\Gamma^{3{\rm d}\,\zeta\zeta\zeta}(0,\vec{p},-\vec{p}) = \frac{12}{5} \frac{f_{\rm NL}}{P_\zeta ( p)}\,.
\ee
To read off the corresponding cubic part of the initial state, simply set $t = t_0$ in~\eqref{split}, with the result:
\be
C_{0,\vec{p},-\vec{p}} =  \frac{12}{5} \frac{f_{\rm NL}}{P_\zeta ( p)} \sim {\cal O}(q^0)\,.
\ee
This manifestly violates the analyticity requirement $C_{\vec{q},\vec{p},-\vec{q}-\vec{p}}\sim {\cal O}(q^2)$. Thus initial local
non-Gaussianities result in $A_{ij}\sim q^0$, which violate the consistency relations as expected.

\subsection{Initial Gaussian Modification}
\label{gaussianinitial}

Next let us focus on the contribution to $\Gamma^{3\rm{d}}$ from the first line of~\eqref{split}. In other words, we focus on purely Gaussian
initial states. For concreteness, consider the family of states that are related to the
Bunch-Davies vacuum by a Bogoliubov transformation (so-called Bogoliubov states). To define such a state one
decomposes the curvature perturbation as
\beq
\zeta_{\vec{q}}(\tau)=a_{\vec{q}}u_{\vec{q}}(\tau)+a^\dagger_{-\vec{q}} u^*_{-\vec{q}}(\tau)\,,
\eeq
where $a^\dagger_{\vec q}$ and $a_{\vec q}$ are the creation and annihilation operators, and $\tau$ is conformal time.
The mode function $u_{\vec{q}}(\tau)$ is a linear combination of positive and negative frequency modes
\beq
u_{\vec{q}}(\tau)=\alpha_{\vec{q}} \zeta^{\rm BD}_q(\tau)+\beta_{\vec{q}} \zeta^{\rm{BD}*}_q(\tau)\,,
\label{modefunction}
\eeq
where the coefficients satisfy the normalization condition $|\alpha_{\vec{q}}|^2-|\beta_{\vec{q}}|^2=1$. 
The Bunch-Davies choice is $\alpha_{\vec{q}} = 1$, $\beta_{\vec{q}} = 0$. In slow-roll approximation, for simplicity, the Bunch-Davies mode function $\zeta^{\rm BD}_q$
is given by the familiar Hankel function\footnote{We also assume $c_s= 1$ for simplicity.}
\beq
\zeta^{\rm BD}_q(\tau) = \frac{H}{\sqrt{2q^3}}(1-iq\tau)e^{iq\tau}\,.
\eeq
By definition, the Bogoliubov vacuum state is annihilated by all $a_{\vec{q}}\,$:
\beq
a_{\vec{q}} |\Omega\ra=0\,, \quad \forall \, \vec{q}.
\eeq
See Appendix~A for the construction of the wavefunctonal corresponding to this state.

As it follows from the explicit form of the mode functions, the non-equal time 2-point function for $\zeta$ is given by
\beq
P_\zeta (q,\tau,0)\propto \frac{1}{q^3}\bigg( \mathcal{A}_{\vec{q}} (1-iq\tau)e^{iq\tau}+\mathcal{B}_{\vec{q}} (1+iq\tau)e^{-iq\tau} \bigg)\,,
\label{bogprop}
\eeq
where $\mathcal{A}_{\vec{q}}$ and $\mathcal{B}_{\vec{q}}$ are related to the Bogoliubov's coefficients by
\beq
\mathcal{A}_{\vec{q}}=(\alpha_{\vec{q}}-\beta_{\vec{q}})\alpha_{-\vec{q}}^*\,,\qquad \mathcal{B}_{\vec{q}}=-(\alpha_{\vec{q}}-\beta_{\vec{q}})\beta_{-\vec{q}}^*\,.
\eeq
The Bunch-Davies vacuum corresponds therefore to $\mathcal{A}_{\vec{q}}=1$ and $\mathcal{B}_{\vec{q}}=0$.

In order to understand how the $q$-dependence is translated in the 3d formalism, let us consider for illustrative purposes the particular interaction term
\beq
S_{\dot{\zeta}^3} \sim \int {\rm d} t {\rm d}^3x\, a^3(t)\dot{\zeta}^3\,.
\label{dzeta3}
\eeq
It is well known that this interaction term in slow-roll inflation behaves as $q^2$ in the soft limit. 
The late-time contribution to $\Gamma^{3\rm{d}}$ from the first line of~\eqref{split}, after converting to conformal time, is given by
\beq
\Gamma^{3{\rm d}\,\zeta\zeta\zeta}(\vec{q},\vec{p},-\vec{q}-\vec{p}) \sim -i\int_{\tau_0}^0{\rm d}\tau'a(\tau')\frac{\p_{\tau'}P_\zeta (q,\tau',0)}{P_\zeta (q)} \, \frac{\p_{\tau'}P_\zeta (p,\tau',0)}{P_\zeta ( p)}\,\frac{\p_{\tau'}P_\zeta (|\vec{q}+\vec{p}|,\tau',0)}{P_\zeta (|\vec{q}+\vec{p}|)} + {\rm c.c.}
\label{bispectrum}
\eeq
From \eqref{bogprop}, each 2-point function in~\eqref{bispectrum} contains both positive and negative frequency contributions, in contrast to the Bunch-Davies vacuum. Based on this observation, it was noticed in~\cite{Agarwal:2012mq} that one gets enhanced non-Gaussianity in the squeezed limit from the cross terms. Specifically, the enhanced contributions are of the form
\beq
\Gamma^{3{\rm d}\,\zeta\zeta\zeta}(\vec{q},\vec{p},-\vec{q}-\vec{p}) \supset i\frac{p^4}{H}\frac{\mathcal{A}_{\vec q}\mathcal{A}_{\vec p}\mathcal{B}_{-\vec{p}-\vec{q}}}{(\mathcal{A}_{\vec q}+\mathcal{B}_{\vec q})(\mathcal{A}_{\vec p}+\mathcal{B}_{\vec p})(\mathcal{A}_{-\vec{p}-\vec{q}}+\mathcal{B}_{-\vec{p}-\vec{q}})}\int_{\tau_0}^0{\rm d}\tau q^2\tau^2 e^{i(q+p-|\vec{p}+\vec{q}|)\tau}+{\rm c.c.}\,~~
\label{aba}
\eeq
Notice that the physical soft limit corresponds to $\vec q\to 0$, $\tau_0\rightarrow -\infty$ keeping $q\tau_0>1$ fixed. In this way, the long mode started out inside the horizon. 
For such a squeezed limit, it is straightforward to see that\footnote{In deriving~\eqref{integral} we assumed $(1 -\hat{q}\cdot \hat{p})q\tau_0\gg 1$, which breaks down
in the collinear limit $\hat{q}\cdot\hat{p}\simeq 1$. Hence the apparent divergence in this limit is an artifact of our approximation.}
\beq
\label{integral}
\lim_{q\to 0\,  \& \, q\tau_0\gg 1}\int_{\tau_0}^0{\rm d}\tau q^2\tau^2 e^{i(q+p-|\vec{p}+\vec{q}|)\tau}\to i\frac{q \tau_0^2}{1-\hat{q}\cdot \hat{p}}e^{i\left(1-\hat{q}\cdot \hat{p}\right)q\tau_0}\,.
\eeq
Hence, as advocated, we see that 
\beq
\lim_{\vec q\to 0}\Gamma^{3{\rm d}\,\zeta\zeta\zeta}(\vec{q},\vec{p},-\vec{q}-\vec{p})\sim q \,.
\label{orderq}
\eeq
This is interesting because a contribution that naively looked irrelevant up to ${\cal O}(q^2)$ becomes ${\cal O}(q)$. This
enhancement traces back to the relative sign between momenta in the phase of the integrand of~\eqref{aba}. As result,
in the case of modified initial states one cannot argue that $A_{ij}\propto q^2$, even if the initial state is Gaussian.

Although we so far focused on the $\dot{\zeta}^3$ interaction, it is straightforward to show that the same conclusion applies to other vertices. In general, the amplitude scales as $q$, but the actual function is
shape-dependent and peaks in the limit of flattened triangle. This can be checked with the exact expression derived in~\cite{Agarwal:2012mq},
\beq
\label{gamma3zeta}
\Gamma^{3{\rm d}\,\zeta\zeta\zeta}(\vec{q},\vec{p},-\vec{q}-\vec{p}) \propto \frac{q^2p^4\sum_{l,m,n=0}^1 c_{\vec q}^{(l)}c_{\vec p}^{(m)}c_{-\vec{p}-\vec{q}}^{(n)} ~\mathcal{F}\left( (-1)^l q,(-1)^m p,(-1)^n |\vec p+\vec q| \right)}{(\mathcal{A}_{\vec q}+\mathcal{B}_{\vec q})(\mathcal{A}_{\vec p}+\mathcal{B}_{\vec p})(\mathcal{A}_{-\vec{p}-\vec{q}}+\mathcal{B}_{-\vec{p}-\vec{q}})}+{\rm c.c.}\,,
\eeq
with $c_{\vec q}^{(0)}=\mathcal{A}_{\vec q}$ and $c_{\vec q}^{(1)}=\mathcal{B}_{\vec q}$. The shape function $\mathcal{F}$ is defined by
\beq
\label{shape}
\mathcal{F}(p_1,p_2,p_3)=-\frac{2}{K^3}+\frac{e^{iK\tau_0}}{K}\left( \frac{2}{K^2}-\frac{2i\tau_0}{K}-\tau_0^2 \right)\,,
\eeq
where $K\equiv p_1+p_2+p_3$. For completeness let us investigate the behavior of~\eqref{gamma3zeta} in different squeezed limits. We focus on the cross-term contribution of the type~\eqref{aba} since
it characterizes the deviation from the Bunch-Davis vacuum. In other words, let us look more closely at the soft limits of $\mathcal{F}_{\cal AAB}$:

\begin{itemize}

\item ${q\tau_0 \ll 1}$. In this case, the mode of interest is so long that it was outside the horizon even at the onset of inflation. Since $K=q+p-|\vec{p}+\vec{q}|\sim q$ for $\mathcal{F}_{\cal AAB}$,
in the limit $q\tau_0 \ll 1$ we can expand in powers of $K\tau_0$:
\beq
\mathcal{F}_{\mathcal{AAB}}=\frac{i\tau_0^3}{3}+\frac{\tau_0^4}{4}K+\mathcal{O}(K^2)\,.
\eeq
Hence, in this extreme limit $\Gamma^{3{\rm d}\,\zeta\zeta\zeta}(\vec{q},\vec{p},-\vec{q}-\vec{p})\sim q^2$, and thus its analyticity property is the same as 
in the Bunch-Davis case. In particular, the consistency relations hold \cite{Kundu:2013gha}. However, it must be stressed
that this regime is not physically interesting since the long mode is always outside the horizon.\footnote{Note that the flattened triangle limit considered earlier belongs to this regime
since $K=q+p-|\vec{p}+\vec{q}|=0$. One would conclude that the $\dot{\zeta}^3$ amplitude should satisfy the consistency relations, in apparent contradiction with~\eqref{integral}.
The resolution is that, the flat limit, the expansion performed in~\eqref{integral} is inapplicable.}

\item ${q\tau_0 \gg 1}$ \textit{\&} $q\ll p$. This is the more physically-interesting case where the long mode started out inside the horizon. Since we have already addressed the
case of flattened triangle, let us now consider $K\sim q\neq 0$. It is straightforward to show that~\eqref{shape} reduces to
\beq
\mathcal{F}_{\mathcal{AAB}}\simeq -\frac{\tau_0^2}{K}e^{iK\tau_0}\,; \qquad  K\simeq q\left(1 -\hat{q}\cdot\hat{p}\right)\,.
\eeq
Hence, $\Gamma^{\rm 3d}\sim q$ in this case, rather than $q^2$, and the consistency relations are violated. In other words, model-dependent contributions to the action, which we expect to be of order $q^2$, contribute to
3d interaction vertices at order $q$ in the case of excited Gaussian initial states. This confirms in generality what we found earlier in~\eqref{orderq}.

\end{itemize}

\section{Conclusion}
\label{conclu}

In this paper, we studied the impact of the initial state on the validity of the inflationary consistency relations, using the fixed-time path integral formalism.
First, we showed that diffeomorphism invariance requires the initial wavefunctional to satisfy a Slavnov-Taylor identity similar to that of the action. Second, we identified two ways in which non-Bunch-Davies initial states
can lead to violations of the consistency relations: $i)$ the initial state can include non-Gaussianities that are not constrained by symmetries; $ii)$ even if the initial state
is Gaussian, such as a Bogoliubov state, the modified non-equal time 2-point functions appearing in the $4{\rm d}\to 3{\rm d}$ translation can map an analytic 
$\Gamma^{4{\rm d}}$ to a non-analytic $\Gamma^{3{\rm d}}$.

As future directions, it would be interesting to study the impact of modified initial states on the consistency relations for the large-scale structure~\cite{Kehagias:2013yd,Peloso:2013zw,Creminelli:2013mca,Horn:2014rta}, as well as those of the conformal alternative to inflation~\cite{Rubakov:2009np,Creminelli:2010ba,LevasseurPerreault:2011mw,Hinterbichler:2011qk,Hinterbichler:2012mv,Creminelli:2012my,Hinterbichler:2012fr,Creminelli:2012qr,Hinterbichler:2012yn,Elder:2013gya}. \\

\noindent {\bf Acknowledgements:} We would like to thank Nishant Agarwal, Richard Holman, Lam Hui, Mehrdad Mirbabayi and Andrew Tolley for useful discussions. L.B. is supported by 
funds provided by the University of Pennsylvania. J.K. is supported in part by NSF CAREER Award PHY-1145525. L.B. is supported by funds from the University of Pennsylvania.

\section*{Appendix A: Wavefunctional for Bogoliubov States}
\renewcommand{\theequation}{A-\Roman{equation}}
\setcounter{equation}{0} 

In this Appendix, we construct explicitly the wavefunctional for Bogoliubov states, {\it i.e.} states related to the Bunch-Davies vacuum by a Bogoliubov transformation. For simplicity, we focus on the curvature perturbation $\zeta$. Its decomposition in terms of the creation and annihilation operators is given by
\beq
\zeta(\vec{x},\tau)=\int{\rm d}^3 p \left( a_{\vec{p}}u_{\vec{p}}(\tau)e^{i\vec{p}\cdot \vec{x}}+a^\dagger_{\vec{p}}u^*_{\vec{p}}(\tau)e^{-i\vec{p}\cdot \vec{x}} \right)\,,
\eeq
where the mode function $u_{\vec{p}}(\tau)$ is given by \eqref{modefunction}. Manifestly, the $a_{\vec{p}}^\dagger$ and $a_{\vec{p}}$ operators are linear combinations of the usual creation and annihilation operators. The conventional vacuum choice is $\beta_{\vec p}=0$ and $\alpha_{\vec p}=1$.

By definition, the Bogoliubov vacuum state is annihilated by $a_{\vec{p}}$ for all momenta. To write down the corresponding wavefunctional, we must express the annihilation operators in terms of (the Fourier components of) the field $\zeta$ and its conjugate momentum $\Pi\equiv \dot{\zeta}$ at some fixed time $\tau_0$:
\beq
a_{\vec{p}}=D_{\vec p} \zeta_{\vec p}(\tau_0)+E_{\vec p}\Pi_{\vec p}(\tau_0) \,,
\eeq
with
\beq
D_{\vec p}= \left[u_{\vec p}(\tau_0)\left( 1-\frac{\dot{u}_{\vec p}(\tau_0)}{u_{\vec p}(\tau_0)}\frac{u^*_{-\vec p}(\tau_0)}{\dot{u}^*_{-\vec p}(\tau_0)} \right)\right]^{-1} \quad \text{and} \quad E_{\vec p}=-\frac{u^*_{-\vec p}(\tau_0)}{\dot{u}^*_{-\vec p}(\tau_0)}E_{\vec p}\,.
\label{ABexpl}
\eeq
In the configuration basis, $\Pi_{\vec p}$ acts as $-i\delta/\delta\zeta_{-\vec p}$ as usual. The vacuum condition $a_{\vec{p}} |\Omega\ra =0$ becomes
\beq
\left( D_{\vec{p}} \zeta_{\vec{p}}(t_0) -iE_{\vec{p}}\frac{\delta}{\delta\zeta_{-\vec{p}}(t_0)} \right)\Psi[\zeta,\tau_0]  =0\,,
\eeq
where $\Psi[\zeta,\tau_0] = \la\zeta\vert \Omega\ra$ is the initial wavefunctional. This solution is a Gaussian state
\beq
\Psi[\zeta,\tau_0] \sim e^{i\mathcal{S}}\,,
\eeq
where
\beq
\mathcal{S}=-\frac{1}{2}\int {\rm d}^3 p \frac{D_{\vec p}}{E_{\vec p}}\zeta_{\vec p} (\tau_0)\zeta_{-\vec p}(\tau_0)\,.
\eeq
Substituting the explicit form for $D$ and $E$ given in~\eqref{ABexpl}, we get
\beq
\mathcal{S}=\frac{1}{2}\int {\rm d}^3 p \left( \frac{p^2\tau_0}{1+ip\tau_0}+\frac{2i\beta_{\vec p}^*p^3\tau_0^2e^{2ip\tau_0}}{\alpha_{\vec p}^*(1+ip\tau_0)^2+\beta_{\vec p}^*(1+p^2\tau_0^2)e^{2ip\tau_0}} \right)\zeta_{\vec p}(\tau_0) \zeta_{-\vec p}(\tau_0)\,.
\eeq
The first term in parantheses brackets corresponds to the vacuum contribution of the standard Bunch-Davies state. The second term parametrizes excitations from that state.

\end{document}